# Thermographic analysis of turbulent non-isothermal water boundary layer

Irina A. Znamenskaya [*], Ekaterina Y. Koroteeva

Faculty of Physics, Lomonosov Moscow State University, Moscow, Russia
*corresponding author: znamen@phys.msu.ru

The paper is devoted to the investigation of the turbulent water boundary layer in the jet mixing flows using high-speed infrared (IR) thermography. Two turbulent mixing processes were studied: a submerged water jet impinging on a flat surface and two intersecting jets in a round disc-shaped vessel. An infrared camera (FLIR Systems SC7700) was focused on the window transparent for IR radiation; it provided high-speed recordings of heat fluxes from a thin water layer close to the window. Temperature versus time curves at different points of water boundary layer near the wall surface were acquired using the IR camera with the recording frequency of 100 Hz. The time of recording varied from 3 till 20 min. The power spectra for the temperature fluctuations at different points on the hot-cold water mixing zone were calculated using the Fast Fourier Transform algorithm. The obtained spectral behavior was compared to the Kolmogorov "-5/3 spectrum" (a direct energy cascade) and the dual-cascade scenario predicted for quasi-2D turbulence (an inverse energy cascade to larger scales and a direct enstrophy cascade to smaller scales).

## 1 Introduction

Infrared (IR) thermography is a non-contact, non-destructive optical tool that allows for visualization and measurement of fully two dimensional temperature maps. Due to its high sensitivity and low response time, this technique has been shown to be advantageous for analyzing complex thermal flows [1, 2].

The IR camera detects the electromagnetic radiation emitted from the studied object and converts it into an instantaneous map of a measured signal. The recent improvement of the IR cameras in terms of their temporal and spatial resolutions made it possible to acquire the thermal patterns from the surface with the rate up to hundreds of frames per second. This allowed for capturing fast changes in the thermal behavior of flows and, hence, for quantitative studies of unsteady thermal processes including flow turbulence [3].

Most of the engineering flow problems, such as thermal mixing, are turbulent in nature. A vast variety of techniques and methods, both numerical and experimental, has been used to address the different aspects of turbulence and turbulent flows. Recently, a new research technique has been proposed to study non-isothermal pulsations of turbulent water boundary layers based on high-speed thermographic measurements through IR-transparent windows [3, 4]. Since a thin layer of liquid water absorbs almost all the IR-radiation, the time-evolving temperature maps of near-wall boundary layers can be recorded by using IR-transparent vessel walls.

In this work, high-speed IR measurements are used to study the thermal pulsations of water boundary layer associated with two non-isothermal mixing processes: submerged hot water jet impinging on a flat vertical surface and two intersecting jets in a round disc-shaped vessel (Fig. 1). Here, the heat serves as a tracer for the turbulence. The power spectra of temperature variations at the selected points on the obtained temperature maps are calculated and compared to the spectral behavior predicted by the theory of 2D and 3D turbulent flows.

## 2 Experimental setup and procedure

The two types of performed experiments are sketched in Fig. 1. In experiments with an impinging jet (Fig. 1, left), the hot water jet was submerged in the vessel with quiescent cold water and injected perpendicularly to the flat vessel wall. The impingement time was 3 minutes. The vessel was a polypropylene container with a wall thickness of 2 mm partially transparent to the IR radiation [3]. In experiments with two intersecting jets (Fig. 1, right), the water was injected via two nozzles, 7 mm in diameter each, set at an angle of 120° to each other. The outflow nozzle, 14 mm in diameter, was located at the upper part of the vessel. The thin disc-shaped vessel was 67 mm in diameter. The vessel window was also made of material, partially transparent in the IR spectral range. The mixing duration was from 5 till 20 min.

The hot and cold water temperatures were in the range of 30-55ºC and 10-20ºC, respectively. The water flow rate through each nozzle varied from 10 to 55 ml/s. The average flow velocity was up to 1,3 m/s, which corresponded to the flow Reynolds number of approximately $1,7 \times 10^4$.

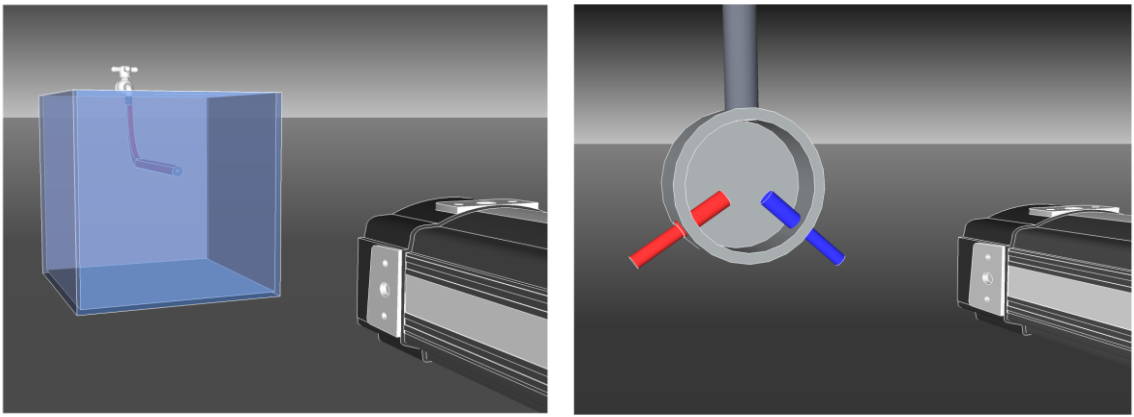

Fig. 1 Sketch of two studied flows: a submerged impinging jet (left); two intersecting jets in a disc-shaped vessel (right).

In all the experiments the time-evolving temperature maps of the water mixing processes were recorded by a calibrated, mid-wave (3,7-4,8 μm) infrared camera (FLIR Systems SC7700). The frame rate was set to 100 Hz, with the frame resolution of 640x512 px. The camera was focused on the inner surface of the IR-transparent vessel walls.

The obtained experimental data (thermal movies) were processed as follows:
1) several characteristic points on each temperature map were selected;
2) the time dependences of measured thermal signals at each point were extracted using the FLIR Altair software;
3) the energy power spectra for each time-series were calculated using the Fast Fourier Transform (FFT) algorithm;
4) the characteristic frequencies of temperature pulsations and the power laws of the obtained power spectra were determined.

It should be noted that the spectral characteristics of the recorded pulsations, unlike their amplitude, were found to be independent of the temperature differences of the mixing jets. Thus, the temperature had a tracer function and acted merely as a passive contaminant (like ink or dye).

## 3 Results and discussion

Turbulent flows are complex, unsteady, and characterized by a wide range of length and time scales. The distribution of the kinetic energy over the hierarchy of scales is the key parameter of any turbulent flow. For isotropic turbulence, it is usually evaluated by means of the energy spectrum function based on a Fourier representation of turbulent fluctuations. The Kolmogorov's theory predicts that the energy density of fully developed isotropic homogeneous turbulence decreases by five decades when the frequency increases by three decades [5,6]. According to this theory, during the turbulent mixing process, energy should "cascade" from large-scale turbulent structures to the smaller ones until it is dissipated by viscous effects.

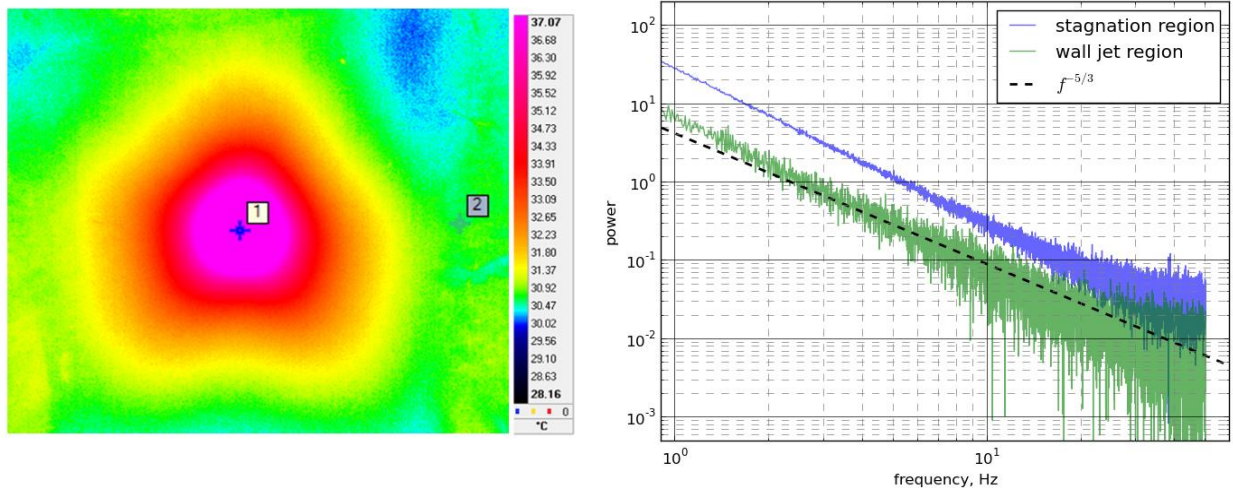

Fig. 2 Infrared image of a submerged jet impinging on an IR-transparent wall (left) and power spectra of thermal pulsations at two selected points: (1) – stagnation region; (2) – wall jet region; dashed black line indicates the standard Kolmogorov power spectrum with an exponent, $\alpha=5/3$.

Turbulence is an intrinsically three-dimensional phenomenon. In certain cases, however, the motion in one direction can be constrained by geometry or applied body forces; this affects qualitatively the dynamics and the spectral properties of turbulent pulsations. The Kraichnan's theory [7] suggests that two-dimensional turbulence can exhibit two inertial ranges: an inverse energy cascade to larger scales (with an energy spectrum scaling of $k^{-5/3}$) and a direct enstrophy (squared vorticity) cascade to smaller scales (with an energy spectrum scaling of $k^{-3}$). A large amount of research has been devoted to the numerical simulations of two-dimensional turbulent flows; experimental work is, however, still limited, since for a long time it has been generally considered that the two-dimensional turbulence cannot be achieved in the laboratory [8,9]. The first experimental observation of an inverse energy cascade can be found in the work by Someria [10], in which the two-dimensionality of the flow is maintained by applying a magnetic field.

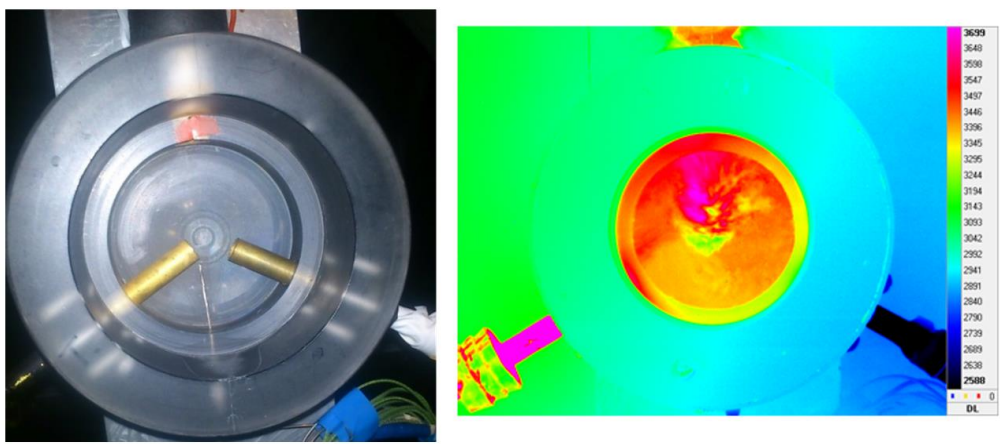

Fig. 3 Photo of a round disc-shaped vessel (left) and infrared image of mixing cold (right pipe) and hot (left pipe) intersecting water jets (right).

In this work, we used the high-speed thermographic measurements to detect the temperature variations during different regimes of non-isothermal turbulent water mixing. To analyze the obtained time-dependant data from thermal images in the frequency domain, we applied the Fourier transform algorithm. According to the Nyquist theorem, the recording rate of 100 Hz allows identifying the temperature fluctuations at frequencies up to 50 Hz.

Fig. 2 shows the power spectra of thermal pulsations of a submerged water jet near the IR-transparent impingement wall at two selected points located in the stagnation and in the wall jet region of the boundary layer. The spectrum clearly follows the usual power law $f^{-\alpha}$, where the spectral index is close to the Kolmogorov value (this spectrum is indicated by a dashed line). This inertial-range scaling of the turbulent energy spectrum is generally observed in three-dimensional fully developed isotropic turbulence [5,7].

The same experimental methodology was used to study the mixing of two intersecting water jets in a round disc-shaped vessel (Fig. 3). Fig. 4 shows the results of the high-speed IR themographic measurements of non-isothermal mixing of two water jets having equal flow rates of 50 ml/s. Three points on the obtained time-evolving temperature map are chosen to illustrate the typical spectral behavior of temperature fluctuations near the water boundary layer. A spectral break at about 8-9 Hz dividing the spectra into two intervals is clearly visible for points 1 and 2. The power laws calculated for each interval using the least squares method (red lines) are also shown.

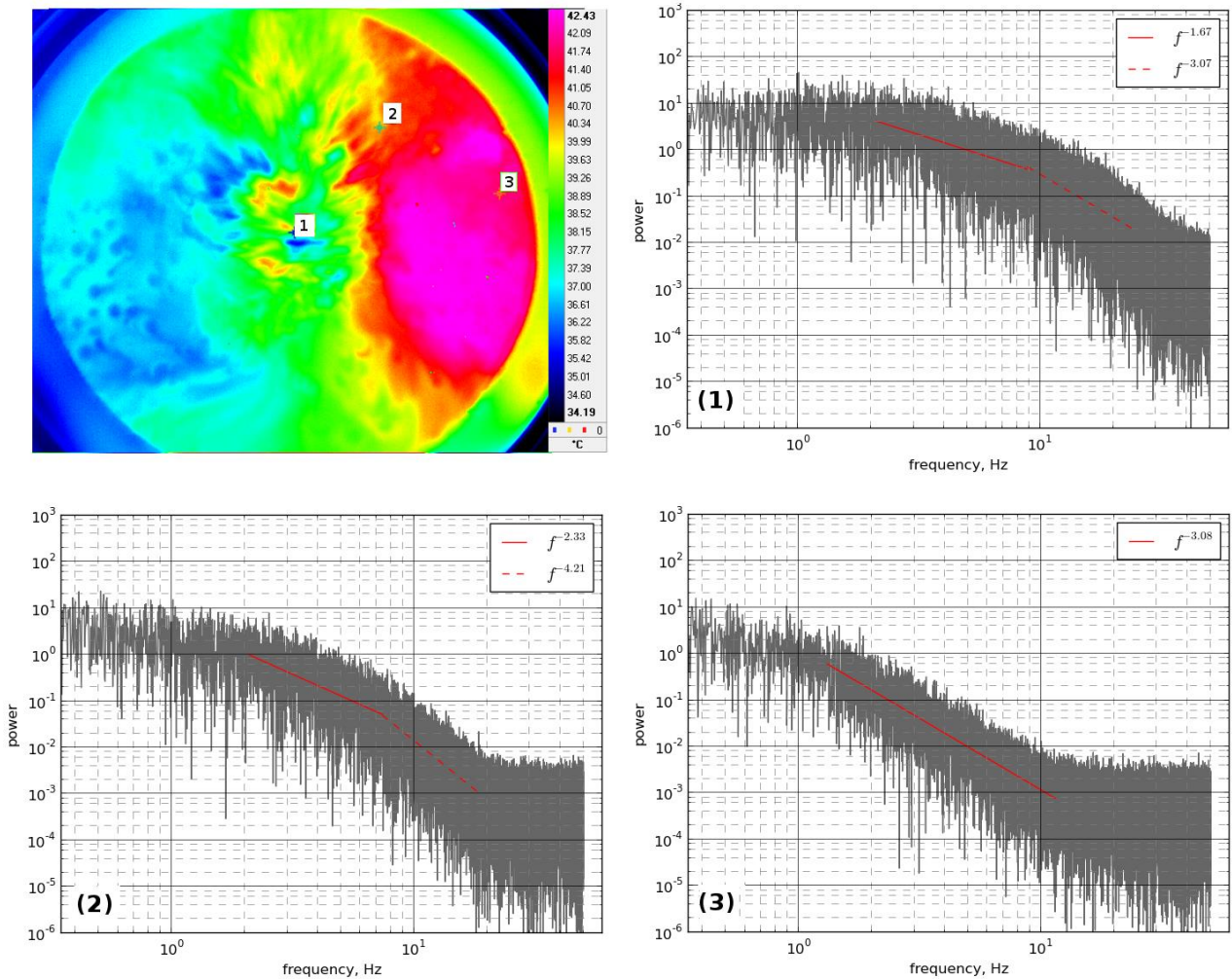

Fig. 4 Power spectra of thermal pulsations at three selected points on the thermal image of two intersecting jets in the round vessel; jets flow rate is 50 ml/s, the recording duration is 600 s.

The analysis of the obtained spectra shows that for points located at the flow centerline (like point 1), the fluctuation spectrum at the low frequency range (1-9 Hz) is consistent with the -5/3 scaling law whereas the second range (9-25 Hz) is consistent with the -3 scaling exponent. The spectral ranges measured using the proposed technique are, accordingly, from 0.5 to 1.5 decades. For higher frequencies, a flattening is usually seen on the spectral plots, which is likely due to the limited experimental resolution. For points located further from the flow centerline (like 2 in Fig. 4), the spectrum of thermal pulsations steepens but it is still described by a power law with a slope in the range $1.7 < α < 2.5$ and $3.0 < α < 4.0$, respectively. Closer to the vessel border, the power spectra steepen significantly, and one inertial range with the spectral index of $α=3.0±0.5$ is usually observed (3 in Fig. 4).

We also performed a series of experiments, in which the ratio of cold and hot jet flow rates varied from 0.23 to 1.0. In case of non-equal flow rates, the flow structure visualized on infrared images was non-symmetrical, and the flow centerline shifted in the direction of the weaker jet. Fig. 5 shows that the value of the average frequency at which the spectral break is observed decreases with decreasing the ratio of the jets flow rate.

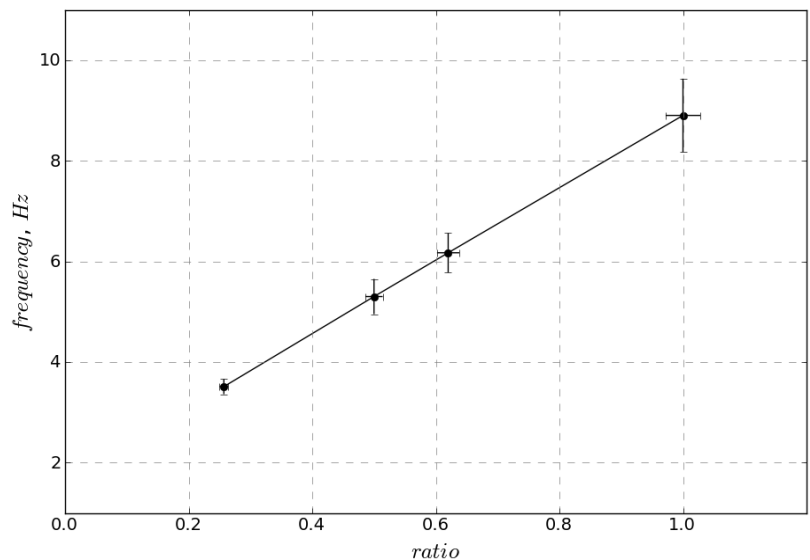

Fig. 5 The average frequency of the spectral break depending on the ratio of the cold and hot water jet flow rates

# 4 Conclusions

High-speed IR thermography was used here to study the thermal pulsations of water boundary layer associated with two non-isothermal mixing processes: submerged hot water jet impinging on a flat vertical surface and two intersecting jets in a round disc-shaped vessel. We calculated the power spectra of temperature variations at the selected points on the obtained temperature maps and compared them to the spectral behavior predicted by the theory of 2D and 3D turbulent flows.

The power spectra of thermal pulsations recorded from the water boundary layer of an impinging jet were found to be consistent with the classical Kolmogorov "-5/3" spectrum. The spectral analysis of the intersecting jets mixing in a round disc-shaped vessel suggests that the coexistence of two scaling regimes typical for quasi-two-dimensional turbulence is possible for such flows (a direct enstrophy cascade at small scales and an inverse energy cascade at large scales).

The obtained results proved that the method based on high-speed IR-thermography can be fruitful for the turbulence analysis. In this case, temperature serves as a passive contaminant and does not affect spectral characteristics of flow pulsations.

# References

[1] Carlomagno G M, Cardone G (2010) Infrared thermography for convective heat transfer measurements. *Experiments in Fluids*, vol. 49(6), pp 1187–1218, http://doi.org/10.1007/s00348-010-0912-2

[2] Carlomagno G M, Ianiro A (2014) Thermo-fluid-dynamics of submerged jets impinging at short nozzle-to-plate distance: A review. *Experimental Thermal and Fluid Science*, vol. 58, pp 15–35, http://doi.org/10.1016/j.expthermflusci.2014.06.010

[3] Znamenskaya I A, Koroteeva E Y (2013) Time-resolved thermography of impinging water jet. *Journal of Flow Visualization and Image Processing*, vol. 20 (1-2), pp 25–33, http://doi.org/10.1615/JFlowVisImageProc.2014010369

[4] Bol'shukhin M A, Znamenskaya I A, Sveshnikov D N, Fomichev V I (2014) Thermographic study of turbulent water pulsations in nonisothermal mixing Optoelectronics, *Instrumentation and Data Processing*, vol. 50 (5), pp 490-497

[5] Kolmogorov A N (1941) Dissipation of Energy in Locally Isotropic Turbulence. *Dokl. Akad. Nauk SSSR*, vol. 32 (1), pp 19–21; and (1991) The local structure of turbulence in incompressible viscous fluid for very large Reynolds numbers, *Proceedings of the Royal Society A*, vol. 434, pp 9–13.

[6] Batchelor G K (1953) The theory of homogeneous turbulence, *Cambridge University Press.*

[7] Kraichnan R H (1967) Inertial Ranges in Two-Dimensional Turbulence. *Physics of Fluids*, vol. 10 (7), pp 1417-1423, http://doi.org/10.1063/1.1762301

[8] Boffetta G, Ecke R E (2012) Two-Dimensional Turbulence. *Annual Review of Fluid Mechanics*, vol. 44 (1), pp 427–451, http://doi.org/10.1146/annurev-fluid-120710-101240

[9] Kellay H, Goldburg W I (2002) Two-dimensional turbulence: a review of some recent experiments. *Reports on Progress in Physics*, vol. 65 (5), pp 845–894, http://doi.org/10.1088/0034-4885/65/5/204

[10] Sommeria J (1986) Experimental study of the two-dimensional inverse energy cascade in a square box. *Journal of Fluid Mechanics*, vol. 170, pp 139-168, http://doi.org/10.1017/S0022112086000836